\PassOptionsToPackage{pdfpagelabels=false}{hyperref} 
\documentclass[fleqn,usenatbib]{mnras}

\usepackage[T1]{fontenc}
\usepackage{ae,aecompl}

\usepackage{natbib}

\usepackage[dvips]{graphics}
\usepackage{epsfig}
\usepackage{amsmath}
\usepackage{aas_macros}

\def\mearth{{\rm\,M_\oplus}}

\title[Outward migration of Jupiter's core]{Did Jupiter's core form in the innermost parts of the Sun's protoplanetary disk?}

\author[Raymond et al]{Sean N. Raymond$^{1,2}$\thanks{E-mail: rayray.sean@gmail.com}, 
Andre Izidoro$^{1,2,3}$,
Bertram Bitsch$^{4}$,
Seth A. Jacobson$^{5,6}$\\
$^{1}$CNRS, Laboratoire d'Astrophysique de Bordeaux, UMR 5804, F-33270, Floirac, France \\
$^{2}$Univ. Bordeaux, Laboratoire d'Astrophysique de Bordeaux, UMR 5804, F-33270, Floirac, France.\\
$^{3}$Capes Foundation, Ministry of Education of Brazil, Bras{\'i}lia/DF 70040-020, Brazil\\
$^{4}$Lund Observatory, Department of Astronomy and Theoretical Physics, Lund University, 22100, Lund, Sweden\\
$^{5}$Laboratoire Lagrange, UMR7293, Universit{\'e} de Nice Sophia-Antipolis, CNRS, Observatoire de la Cote d'Azur, \\Boulevard de l'Observatoire, 06304 Nice Cedex 4, France\\
$^{6}$Bayerisches Geoinstitut, UniverstŠt Bayreuth, 95440 Bayreuth, Germany
}

\begin{document}

\date{Submitted to MNRAS Nov 25, 2015; Accepted Feb 15, 2016}

\pagerange{\pageref{firstpage}--\pageref{lastpage}} \pubyear{2016}

\maketitle

\label{firstpage}

\begin{abstract}
Jupiter's core is generally assumed to have formed beyond the snow line.  Here we consider an alternative scenario, that Jupiter's core may have accumulated in the innermost parts of the protoplanetary disk.  A growing body of research suggests that small particles (``pebbles'') continually drift inward through the disk.  If a fraction of drifting pebbles is trapped at the inner edge of the disk a several Earth-mass core can quickly grow.  Subsequently, the core may migrate outward beyond the snow line via planet-disk interactions.  Of course, to reach the outer Solar System Jupiter's core must traverse the terrestrial planet-forming region.  

We use N-body simulations including synthetic forces from an underlying gaseous disk to study how the outward migration of Jupiter's core sculpts the terrestrial zone. If the outward migration is fast ($\tau_{mig} \sim 10^4$ years), the core simply migrates past resident planetesimals and planetary embryos.  However, if its migration is slower ($\tau_{mig} \sim 10^5$ years) the core removes solids from the inner disk by shepherding objects in mean motion resonances. In many cases the disk interior to 0.5-1 AU is cleared of embryos and most planetesimals. By generating a mass deficit close to the Sun, the outward migration of Jupiter's core may thus explain the absence of terrestrial planets closer than Mercury. Jupiter's migrating core often stimulates the growth of another large ($\sim$ Earth-mass) core -- that may provide a seed for Saturn's core -- trapped in exterior resonance.  The migrating core also may transport a fraction of terrestrial planetesimals, such as the putative parent bodies of iron meteorites, to the asteroid belt.  
% and has the potential to explain other aspects of the inner Solar System.  }

\end{abstract}


\section{Introduction}

The core accretion model proposes that giant planets form in two steps~\citep{pollack96,ida04}.  First, several-Earth-mass cores grow from the solid (condensable) component of the disk.  These cores subsequently accrete thick envelopes from the gaseous component of the disk.  Thus, giant planet cores are constrained to form within the lifetime of their parent gaseous disk.  Observations of young stars in clusters with a spread of ages suggest that gaseous protoplanetary disks last a few million years~\citep{haisch01,hillenbrand08b}, although given observational biases it is possible that many disks last significantly longer~\citep{pfalzner14}.  

Numerical simulations have shown that the growth of giant planet cores by accreting planetesimals is inefficient and slow~\citep{levison10}.  In contrast, pebble accretion generates much faster core growth~\citep{lambrechts12,morby12}.  A growing core can efficiently accrete bodies for which the stopping time due to aerodynamic gas drag is comparable to the orbital time; these bodies are generally called ``pebbles'' although they can range from millimeters to meters in size depending on the local disk conditions~\citep{weidenschilling77b}.  Models of pebble accretion show that giant planet cores can grow on Myr or shorter timescales~\citep{lambrechts14,levison15,bitsch15b}.  

Because of their coupling with the gas disk, pebbles rapidly drift inward~\citep{weidenschilling77b}.  It is unclear whether these pebbles simply evaporate in the hot inner disk, or whether they can accumulate and grow~\citep{boley14,li15}.  They can in principle be trapped at a pressure bump associated with the inner edge of an MRI-inactive dead zone, and \cite{chatterjee14} have proposed this as a potential building site for hot super-Earths.  Whatever the exact stopping mechanism, a large core would grow rapidly if it could gather even just a few percent of the flux of inward-drifting pebbles.

If Jupiter's core formed by this type of process in the very inner parts of disk, it must have migrated outward to at least several AU.  Large-scale outward migration is indeed in line with current hydrodynamical models~\citep{paardekooper11,bitsch15}. Figure~1 shows three snapshots in the evolution of the disk model from \cite{bitsch15}.  The contours outline regions of outward migration, and the right-hand side of each contour is a zero-torque location.  In each snapshot a several $\mearth$ core would indeed migrate outward from the inner disk.  In the two later snapshots (1 and 2 Myr) a few~$\mearth$ core would migrate out to 3-4 AU.  In the 0.2 Myr snapshot there is a gap between two different outward migration regions that extends from roughly 1 to 3 AU.  However, dynamical torques acting on {\it migrating} planets can propel planets far beyond the zero-torque locations~\citep{paardekooper14,pierens15} and may in some cases bridge this gap.  An additional torque driven by the heat of accretion could also help sustain outward migration as long as $M_p < 5 \mearth$~\citep{benitez15}.

%\begin{figure}
%  \begin{center} \leavevmode \epsfxsize=8.5cm\epsfbox{migration_map_color.eps}
%    \caption[]{type I migration map for the disk model of \cite{bitsch15}.  The contours enclose a region of outward migration.  In this model a few Earth-mass core would migrate from 0.1 AU out to 3 AU.  } 
%     \label{fig:migmap}
%    \end{center}
%\end{figure}

Would the long-distance outward migration of Jupiter's core be consistent with the orbital architecture of the inner Solar System?  There is very little mass interior to Venus' orbit.  Indeed, the terrestrial planets are consistent with having accreted from a narrow annulus of solids initially extending from 0.7 to 1 AU~\citep{hansen09}.  In the Grand Tack model the outer edge of this annulus is attributed to dynamical sculpting the migrating (fully-grown) Jupiter~\citep{walsh11}.  While several models exist to explain the lack of planets closer-in than Mercury (see discussion in section 3.1), the inner edge remains unsatisfactorily explained. 
%Explanations that rely on planet migration do not explain why the Solar System does not contain any super-Earths piled up at the inner edge of the gas disk (Ida & Lin, 2008; Batygin et al. 2014). Questions remain as to whether explanations that wish to grind away inner material are erosive enough to reduce every inner embryo to dust (Volk & Gladman, 2014).

\begin{figure}
%  \begin{center} 
  \leavevmode \epsfxsize=8.5cm\epsfbox{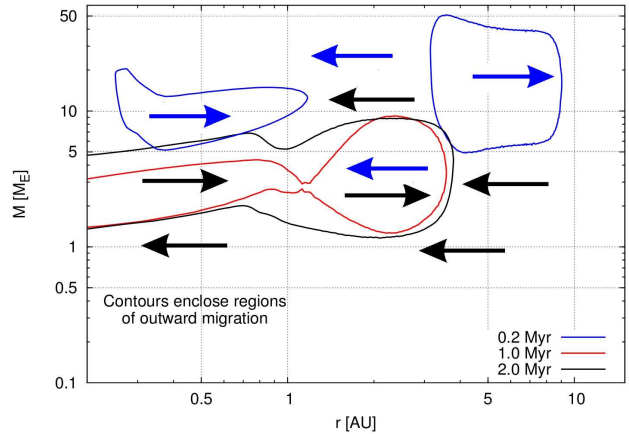}
    \caption[]{{A migration map of the disk model of \cite{bitsch15}.  Each color shows the map at a specific snapshot in time during the disk's evolution. A planet inside the contours migrates outward (to the right in the plot). Arrows indicate the direction of migration at a given snapshot in time: the blue arrows correspond to the 0.2 Myr snapshot and the black arrows to the other two snapshots.  The disc has a metallicity in dust particles of $Z=\Sigma_{\rm Z} / \Sigma_{\rm gas} = 0.5\%$, assumed to be independent of the orbital distance. At the water ice line, water ice grains sublimate and do not contribute to opacity, which is the ultimate cause of the different regions of outward migration~\citep{bitsch14}. The disc features an alpha viscosity of $\alpha=0.0054$~\citep{shakura73}.} }
     \label{fig:migmap}
%    \end{center}
\end{figure}

Could an outward-migrating core explain the Solar System's mass deficit close-in?  When a planet migrates through a disk of smaller bodies, these bodies can be accreted onto the planet, shepherded (essentially pushed) by mean motion resonances, or scattered behind the planet~\citep{zhou05,fogg05,mandell07}.  Close to the star, where the gas surface density is generally higher and the Sun's potential well is deep, shepherding and accretion are favored over scattering~\citep{tanaka99}.  Thus, an outward-migrating core might be effective at clearing out small bodies close to the Sun.  But beyond a given orbital radius, clearing out should become less efficient.  As the surface density drops with orbital radius, shepherding becomes less effective and scattering more important~\citep{tanaka99}, generating instabilities in which shepherded bodies cross the orbit of the migrating core.  The outcome of these instabilities depends on the orbital radius or, more precisely, on the ratio of the escape speed from the core to the local escape speed from the Sun: collisions are favored for closer-in orbits whereas scattering is favored for more distant orbits~\citep{ford08}.  As the core migrates away, scattered objects are simply left behind.

In section 2 we use a suite of dynamical simulations to show that the mechanism of clearing-out of the inner disk by Jupiter's outward-migrating does indeed operate.  In addition, an outward-migrating core can transport inner disk planetesimals outward, possibly implanting the parent bodies of iron meteorites in the asteroid belt. While our simulations are crude they show that, in principle, the outward migration of Jupiter's core could have sculpted the inner Solar System.  While this only holds for a limited range of parameter space, it is a reasonable range.  In section 3 we discuss other existing models to explain the lack of very close-in terrestrial planets.  We then discuss implications of the outward migration of Jupiter's core for terrestrial planet formation, and a simple exploration of a global view of outward-migrating cores. 

\section{Simulations}

We performed a suite of dynamical simulations to test whether it is plausible for Jupiter's core to have formed interior to the terrestrial planets. To accomplish this we simulated the effect of an outward-migrating core (of a few Earth masses) on the building blocks of the terrestrial planets.   In section 2.1 we describe the setup and details of the simulations.  In section 2.2 we outline the metrics we use to determine whether a simulation was successful, and in section 2.3 we present the results of the simulations.  

\subsection{Setup}

Our simulations are assumed to start after Jupiter's core grew to several Earth masses near the inner edge of the gaseous disk. At the start of the simulations, Jupiter's core was taken to be $3 \mearth$ and placed at 0.1 AU, the assumed inner edge of the disk.  A population of terrestrial building blocks consisted of $3 \mearth$ divided equally between 20 planetary embryos and 400 or 1000 planetesimals.\footnote{These initial conditions were inspired by simulations of terrestrial planet formation~\citep[see ][for recent reviews]{morby12b,raymond14}.  A fully self-consistent model would first form planetesimals via the streaming instability, then grow embryos throughout the inner disk using the same pebble flux model for the accreting embryos and Jupiter's core~\citep[see][for a description of some of the relevant steps in this process]{johansen07,lambrechts12,johansen14,morby15a}.} The distribution of embryos and planetesimals started at 0.15 AU and extended out to either 1 AU in some simulations and 5 AU in others.  Embryos and planetesimals were initially evenly spread in orbital radius with small randomly-chosen eccentricities of up to 0.02 and inclinations of up to $1^\circ$. 

These initial conditions were motivated by empirical constraints.  The terrestrial planets comprise a total of $\sim 2 \mearth$ in solids, centered at 0.7-1 AU.  The masses and orbits of the terrestrial planets can be reproduced if $2 \mearth$ in embryos and planetesimals was spread in an annulus between roughly 0.7 and 1 AU during late-stage accretion~\citep{chambers01,hansen09}.  While the current study is focused on the origin of the inner edge of this annulus, the outer edge at 1 AU is essential for reproducing the large Earth/Mars mass ratio. Our two sets of initial conditions essentially represent different assumptions for the origin of the outer edge. The disk that extends only to 1 AU assumes that the outer edge was sculpted by an early, presumably hydrodynamical mechanism that concentrated solids inside 1 AU.  In contrast, the disk that extends to 5 AU assumes that the outer edge of the annulus was sculpted at a later time, possibly via the Grand Tack model during Jupiter's inward-then-outward migration~\citep{walsh11,pierens11}.  In the Grand Tack model, Jupiter's inward migration shepherded material from 1-3 AU inward such that only a few Earth masses are required interior to Jupiter's initial orbit at 3-4 AU~\citep{walsh11}. Thus, our two sets of initial conditions represent different pathways for the formation of the inner Solar System. In the results section below we present the outcome of both sets. We do focus on the simulations in which planetesimals and embryos are only present out to 1 AU simply because they are more convenient to illustrate certain mechanisms at play.

We calculated synthetic forces on all particles in the simulation from an underlying gaseous protoplanetary disk.  We followed the implementation of \cite{mandell07}, based on \cite{thommes03}.  The disk was given a radial surface density profile $\Sigma(r) = 5000 \left(r/1AU\right)^{-1} g \, cm^{-2}$.  We tested both a disk with a constant vertical scale height $H/r = 0.05$~\citep[consistent with the inner parts of the disk in][at the early disk evolution stages]{bitsch15} and a flared disk with $H/r = 0.05 \left(r/1 AU\right)^{1/4}$.  Planetesimal particles felt aerodynamic gas drag~\citep{adachi76} in the form of Stokes drag:
\begin{equation}
{\bf a}_{drag} = - K v_{rel}{\bf v_{rel}},
\end{equation}
where $v_{rel}$ is the relative velocity between a particle and the local gas.  The drag parameter $K$ is defined as:
\begin{equation}
K = \frac{3 \rho_{gas} C_D}{8 \rho_m r_m},
\end{equation}
where $\rho_{gas}$ is the local gas density, $\rho_m$ is the particle's physical density (assumed to be 1 $g 
\, cm^{-3}$ for all planetesimals and 3 $g \, cm^{-3}$ for all embryos), $r_m$ is the particle's radius, and $C_D$ is set to 1~\citep{adachi76}.  The gas is assumed to follow a circular orbit at sub-Keplerian velocity because of internal pressure support, such that 
\begin{equation}
v_{gas} = v_{Kepler} \left(1 - \eta \right),
\end{equation}
where $\eta$ is 
\begin{equation}
\eta = \frac{\pi}{16}\left(\gamma + \delta\right)\left(\frac{H}{r}\right)^2,
\end{equation}
where $\gamma$ and $\delta$ are the disk's radial density and temperature gradients, respectively~\citep[see][]{thommes03}.  In practice $\eta \sim 10^{-3}$ in most of the disk.  Consequently, even planetesimals on circular orbits felt a headwind and spiraled inward. We gave planetesimal particles an effective radius for gas drag and tested values of 100km and 10km.

Embryos and Jupiter's core felt type I damping~\citep{tanaka04,cresswell08} from the disk.  This was calculated by first defining a characteristic timescale $t_{wave}$:
\begin{equation}
t_{wave} = \left(\frac{M_\star}{m}\right)\left(\frac{M_\star}{\Sigma a^2}\right) \left(\frac{H}{r}\right)^{4} \Omega^{-1},
\end{equation}
$M_\star$ is the stellar mass, $m$ is the core mass, $\Sigma$ is the local surface density, $a$ is the semimajor axis and $\Omega$ is the local Keplerian frequency.  Eccentricity $e$ and inclination $I$ damping timescales were calculated following \cite{cresswell08}:
\begin{equation} 
t_e = \frac{t_{wave}}{0.78} \left[1-0.14\left(\frac{e}{H/r}\right)^2 + 0.06 \left(\frac{e}{H/r}\right)^3+0.18\left(\frac{e}{H/r}\right)\left(\frac{I}{H/r}\right)^2\right]
\end{equation}
and 
\begin{equation} 
t_i = \frac{t_{wave}}{0.544} \left[1-0.3\left(\frac{I}{H/r}\right)^2 + 0.24\left(\frac{I}{H/r}\right)^3+0.14\left(\frac{e}{H/r}\right)^2\left(\frac{I}{H/r}\right)\right].
\end{equation}
Finally, accelerations were implemented as follows~\citep[again following][]{cresswell08}:
\begin{equation}
{\mathbf a_e} = -2\frac{\mathbf v \cdot r}{r^2 t_e} {\mathbf r}
\end{equation}
and
\begin{equation}
{\mathbf a_i} = - \frac{\mathbf v_z}{t_i}{\mathbf k},
\end{equation}
where $\mathbf{k}$ is the unit vector in the $z$ direction.

Jupiter's core was forced to migrate outward on a timescale related to the local isothermal type I migration timescale~\citep{ward97,tanaka02}, defined as 
\begin{equation}
\tau_{iso} = \left(\frac{2}{2.7+1.1 \beta}\right) \left(\frac{M_\star}{m}\right)  \left(\frac{M_\star}{\Sigma \, a^2}\right)  \left(\frac{H}{r}\right)^{2} \Omega^{-1}, 
\end{equation}
where $\beta$ is the local surface density profile index (here, $\beta=1$ throughout the disk; note that $\tau_{iso}$ is closely related to $t_{wave}$ -- see \citet{cresswell08}). For simplicity, when calculating the migration timescale we neglected an additional term that accounts for a reduction in migration speed for planets with non-negligible orbital eccentricity. For our chosen disk, $\tau_{iso} \approx 10^4$~years. Given that the outward migration timescale is very hard to estimate reliably~\citep[e.g.,][]{lega14,lega15,fung15}, we tested a range of values: $\tau_{mig} =$ 1, 3, 5, 10 and 30 $\tau_{iso}$. Our constraints on the outward migration timescale were essentially empirical; we did not test timescales shorter than $\tau_{iso}$ or longer than 30 $\tau_{iso}$ simply because preliminary simulations with very short or very long migration timescales failed to match our criteria for success (see Section 2.2 below). The core's outward migration was stopped when it reached 5 AU.    

Our code is based on that of \cite{raymond06c} and \cite{mandell07}, built on the Mercury hybrid integrator~\citep{chambers99}.  Each simulation was evolved for 1 Myr with a 2 day timestep, sufficiently short to avoid numerical errors for orbits as close-in as $\sim 0.05$ AU~\citep[][]{rauch99,levison00,raymond11}.  The disk did not evolve during the simulation.

Our simulations are admittedly crude.  The main effect that is missing is a self-consistent description of the core's migration, which we did not attempt because of the complexity of the problem and the inherent uncertainties.  Modeling the heating torque~\citep{benitez15} would require a feedback between the core's accretion and migration rates.  Meanwhile, to capture the effect of the dynamical corotation torque~\citep{paardekooper14,pierens15} would require keeping track of the history of the local thermal structure during the core's migration.  Finally, the fluid flow in the vicinity of the core is so complex~\citep{lega14,fung15} that we do not feel adequately equipped to calculate a self-consistent migration rate.  

\subsection{Metrics for success}

To evaluate the outcome of our simulations we considered two metrics for success. The first metric is that a successful simulation should clear embryos and planetesimals from the innermost parts of the disk. We have invoked Jupiter's outward-migrating core to explain the absence of terrestrial planets closer-in than Mercury. This metric offers a concrete improvement over standard models of terrestrial planet formation, which empirically invoke an ad-hoc depletion of solids interior to 0.5-0.7 AU to reproduce Mercury's small mass~\citep[e.g.,][]{chambers01,hansen09}.  A simulation that fails to meet this criterion offers no advantage over simpler models and is thus rejected by Occam's razor.  Our second metric is simply that the terrestrial planet-forming zone cannot be entirely depleted.  While the first metric requires depletion of the disk inside $\sim$0.5 AU, the second metric requires that the depletion be confined.  This is of course essential in order to keep enough mass in the terrestrial planet-forming region to reproduce Earth and Venus.  

As we discuss below, some simulations had other successes. In many simulations Jupiter's outward-migrating core stimulated the growth of a second large core in an exterior orbit, locked in outer mean motion resonance.  In addition, in some simulations a group of planetesimals was transported from the inner disk out to the present-day asteroid belt; these transported planetesimals may be interpreted as the parent bodies of iron meteorites~\citep[see][]{bottke06}.  While these outcomes are definite plusses, we do not use them as a metric of success, since other mechanisms may plausibly reproduce them. 

\subsection{Results}

Figure~2 shows three example simulations that illustrate the diversity of outcomes.  The three simulations have a similar setup: they each started from an annulus extending from 0.15 to 1 AU.  The main difference between the simulations is the migration speed of Jupiter's core: $\tau_{mig} = 1, 5$ and $10 \, \tau_{iso}$ (top to bottom; note also that the top two simulations were in a flared disk and contained 20 embryos while the bottom was in a non-flared disk and contained 10 embryos).  Only one of the simulations -- the middle one -- is consistent with the present-day Solar System.  

\begin{figure}
  \begin{center} \leavevmode 
 \epsfxsize=8.75cm \epsfbox{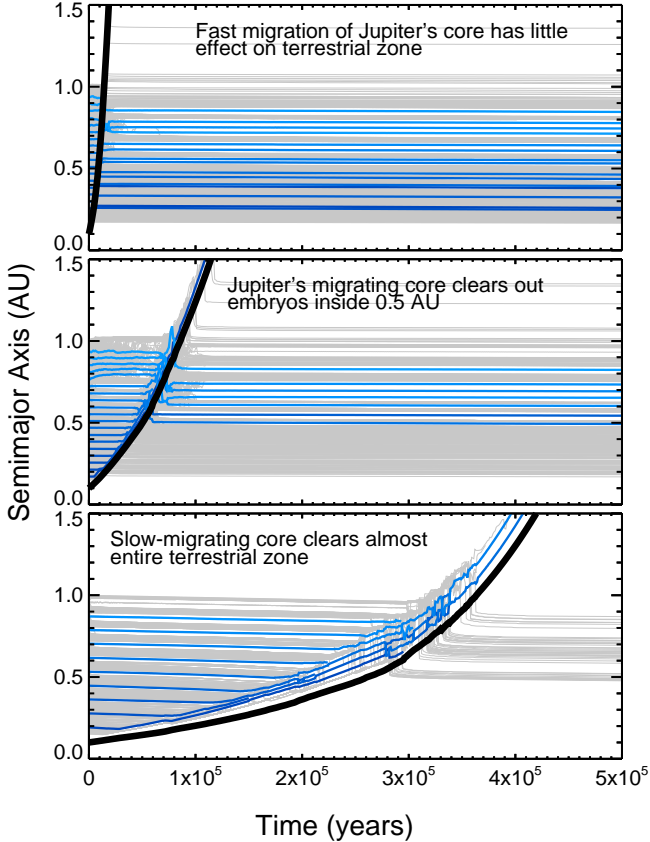}  
    \caption[]{An illustration of the variety of outcomes in our simulations.  Each panel shows the orbital evolution of a simulation that started with an annulus of rocky material extending from 0.15 to 1 AU.  Planetesimals are shown in grey and embryos in shades of blue.  The thick black line represents the migrating core. In the top/middle/bottom simulation, $\tau_{mig} = 1/5/10 \tau_{iso}$, planetesimals were assumed to be 100km/100km/10km in size for the gas drag calculation, and there were 20/20/10 embryos of $0.075/0.075/0.15 \mearth$ each. } 
     \label{fig:at_all}
    \end{center}
\end{figure}

In the top simulation with the fast-migrating core, only 13\% of the mass inside 0.5 AU was cleared out.  The innermost surviving embryo is at 0.22 AU, less than 0.1 AU from its initial location. In the middle simulation the core's migration was 5 times slower and the clearing much more effective.  Three quarters of the mass inside 0.5 AU was cleared out and the innermost surviving embryo was located at 0.5 AU.  The bottom panel shows a simulation in which the entire terrestrial region was cleared of embryos.  Of the $3 \mearth$ initially located in the inner disk, only $0.2 \mearth$ remained in the form of planetesimals.  In this simulation two large embryos ($1.5 \mearth$ and $0.5 \mearth$) accreted efficiently and survived on orbits exterior to Jupiter's core's in a resonant chain.  This was a common feature of simulations with efficient shepherding that we discuss in more detail below.

A migrating core clears out close-in embryos and planetesimals by shepherding -- essentially pushing -- them in exterior resonances.  This is analogous to the mechanism proposed to form large rocky planets close to their stars from material shepherded by migrating gas giants~\citep{fogg05,zhou05,raymond06c,mandell07}.  In test simulations containing only planetesimals, an outward-migrating core removed 100\% of the material inside 1 AU for all of the tested migration rates.  Likewise, a single embryo is cleanly shepherded by a migrating core.  However, shepherding is less efficient when many embryos are present.  Gravitational perturbations between embryos during shepherding can kick them onto core-crossing orbits.  This can result in an embryo-embryo or embryo-core collision or a close encounter with the core. Such close encounters can lead to gravitational scattering or collisions.  In a pure gravitational setting, collisions are favored close to the star, where the escape speed from the star is larger than the escape speed from the planet~\citep[e.g.][]{ford08,raymond10}.  For our $3 \mearth$ core collisions are favored throughout the inner disk; however, the situation is complicated by the fact that the core is migrating away so there is a limited time for a collision to happen.  Thus, whether a collision happens also depends on the core's migration speed. When an embryo is scattered by the core, the embryo may wind up on an orbit interior to the core's~\citep[see][for a description of the same mechanism in a different context]{izidoro14}.  As the core migrates away the embryo can be stranded.  An embryo is more likely to be left behind by this process if the core is migrating quickly and if there are many embryos providing perturbations.  

We thus expect the mechanism of clearing out of the inner disk by an outward-migrating core to be more efficient in situations with few embryos and when the core migrates slowly.  That is indeed what our simulations showed.  The most efficient clearing occurred in simulations in which few embryos were present in the inner disk and Jupiter's core migrated slowly.  Clearing was modestly more efficient in simulations with flared disks.  In flared disks the damping in the inner disk from drag was stronger but the migration time faster; the net result was a modest increase in the efficiency of shepherding.  We saw no systematic difference between the simulations with 10~km and 100~km planetesimals.  

%\begin{figure}
%  \begin{center} \leavevmode 
%%  \epsfxsize=8.5cm\epsfbox{at_zoom_annulus_smallemb_100km_slow.eps}
%%  \epsfxsize=8.5cm\epsfbox{at_zoom_annulus_100km_slowsl.eps}
%%  \epsfxsize=8.5cm\epsfbox{at_zoom_annulus_bigemb_10km_slowslow.eps}  
% \epsfxsize=8.5cm\epsfbox{at_2ex.eps}  
%    \caption[]{Orbital evolution of two simulations that started with an annulus of rocky material extending from 0.15 to 1 AU. Planetesimals are shown in grey and embryos in black.  The thick black line represents the migrating $3 \mearth$ core. In the top/bottom simulation, $\tau_{mig} = 1/10 \tau_{iso}$,  Planetesimals were assumed to be 100km in size for the gas drag calculation, and there were 20 embryos of $0.075 \mearth$ each. } 
%     \label{fig:at_all}
%    \end{center}
%\end{figure}

\begin{figure}
  \begin{center} \leavevmode 
 \epsfxsize=8.75cm\epsfbox{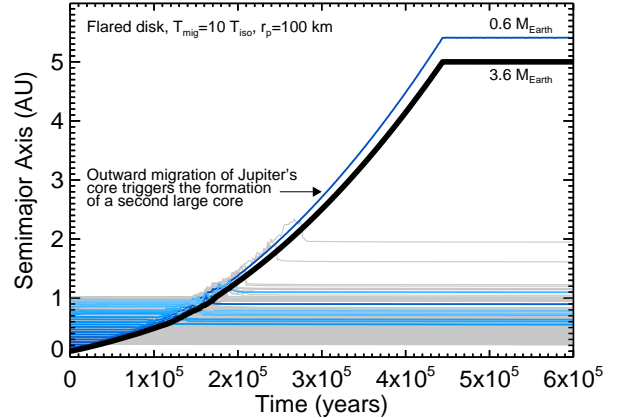}  
    \caption[]{A simulation in which Jupiter's slowly-migrating ($\tau_{mig} = 10 \tau_{iso}$) core cleared out the majority of the mass close to the Sun. It also stimulated the growth of an additional large ($0.6 \mearth$) core in an exterior orbit, which survived in a high-order (9:10) mean motion resonance with Jupiter's core, which also accreted $\sim 0.6 \mearth$.  Planetesimals are shown in grey and embryos in shades of blue.  The disk was flared. } 
     \label{fig:at_mega}
    \end{center}
\end{figure}

In many simulations the outward-migrating core stimulated the growth of one or more additional large cores exterior to its orbit.  This happened in most of the simulations in which the inner disk was efficiently cleared out.  This makes sense since both processes are linked to the efficiency of shepherding.  Figure~\ref{fig:at_mega} shows the evolution of one simulation in which the inner disk was cleared out and a large ($0.6 \mearth$) core accreted and survived just exterior to Jupiter's core, in 9:10 mean motion resonance.  Large cores such as this formed with masses up to $1.5 \mearth$ (the bottom simulation from Fig.~2).  These shepherded cores may serve as seeds for the cores of other gas giants such as Saturn.  

Jupiter's core also grows during migration~\citep[see][]{tanaka99}.  As for shepherding, accretion onto the core is more pronounced for slower migration rates.  In simulations with efficient clearing the core typically accreted $0.2-0.8 \mearth$.  

The evolution of our simulations with broad disks of planetesimals and embryos (extending from 0.15 to 5 AU) is broadly similar to the simulations with narrow disks of solids (from 0.15 to 1 AU).  The most important difference between the broad and narrow disks is the initial spacing of embryos.  Embryos were much more widely-spaced in the simulations with broad disks, with just a handful present in the inner disk.  As expected from the discussion above, the wide spacing of embryos increased the efficiency of shepherding close-in.  

\begin{figure}
  \begin{center} \leavevmode 
 \epsfxsize=8.75cm\epsfbox{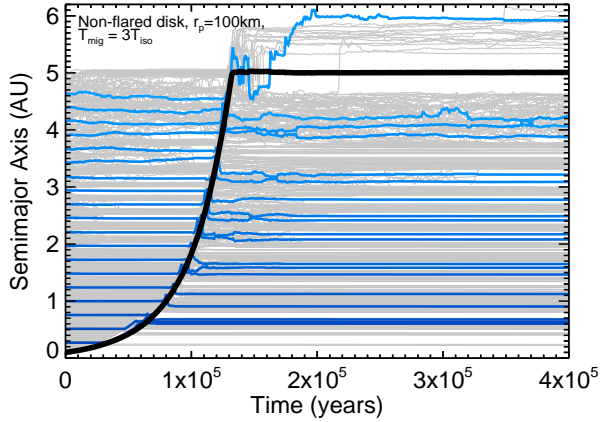}  
    \caption[]{Evolution of a simulation in which embryos and planetesimals were initially distributed from 0.15 to 5 AU.  In this case, Jupiter's migrating ($\tau_{mig} = 3 \tau_{iso}$) core cleared out the bulk of the mass interior to 0.6 AU.  The core had little effect on the distribution of embryos and planetesimals beyond 1 AU. A single small ($0.075 \mearth$) embryo survived on an exterior orbit.  In this simulation the disk was not flared and planetesimals were 100~km in size.
} 
     \label{fig:at_disk}
    \end{center}
\end{figure}

Figure~\ref{fig:at_disk} shows the evolution of a simulation with a broad disk of planetesimals and planetary embryos.  A relatively fast-migrating ($\tau_{mig} = 3 \tau_{iso}$) core cleared out embryos interior to 0.6 AU but did not substantially deplete or excite the region beyond 1 AU.  One embryo and a handful of planetesimals survived exterior to 5 AU.  The only strong effect of the migrating core was to clear out the innermost parts of the disk. 

The simulation from Fig.~\ref{fig:at_disk} was characteristic of the simulations in broad disks. The innermost disk was generally cleared out more efficiently than in the simulations in narrow disks because of the wide embryo spacing.  Whereas, the presence of embryos at several AU often acted to disrupt the shepherding of large embryos by Jupiter's migrating core.  Several simulations produced large outer cores but the efficiency was substantially lower than in the simulations with an annulus, as was the typical outer core mass.  Because of the presence of embryos in the asteroid region, strong effects of the migrating core were confined to the innermost parts of the disk.

Of course, the initial conditions for terrestrial planet formation are poorly constrained.  It is unclear whether embryos ever existed in the primordial asteroid belt. Traditional models of runaway and oligarchic accretion of planetesimals find a strong radial dependence of embryo growth rate~\citep[e.g.][]{kokubo00}.  Yet pebble accretion can form large cores quickly at a range of orbital radii if the conditions are right~\citep{lambrechts12,lambrechts14}.  Our simulations bracket the range of expected situations, from an empty primordial asteroid belt to one containing many large embryos.  What is most important for this study is understanding the physical mechanisms at play and the trends with system parameters, which are clear: many embryos in the asteroid region reduce the efficiency of forming Saturn's core via shepherding.  Likewise, a very high density of embryos close-in reduces the efficiency of clearing out the very inner parts of the disk.

\section{discussion}

\subsection{Other models for the lack of planets in the very inner Solar System}

Other ideas exist to explain the lack of terrestrial planets close to the Sun.  \cite{ida08} proposed that, if embryos grew from planetesimals as an outward-propagating wavefront~\citep[as in the model of][]{kokubo00} and these embryos systematically type I migrated inward and were destroyed, then all embryos within a given orbital radius corresponding to the wave front's edge at the time of disk dispersal could be removed.  However, this idea is based on isothermal migration models and assumes that inward-migrating embryos simply fall onto the Sun, which conflicts with models of magnetized stellar accretion that predict an inner edge to the gas disk creating an uncrossable gap for a migrating planet~\citep[e.g.][]{bouvier07}.  

\cite{morby15b} proposed that the inner edge of \cite{hansen09}'s proposed annulus of embryos represents the silicate condensation front that was ``fossilized'' early in the disk's history.  If planetesimals only formed at early times, perhaps via the streaming instability~\citep{youdin05,johansen07b}, then inward-drifting pebbles might only find seeds on which to grow exterior to the fossilized condensation front.  The main uncertainty in this model is the fate of the pebbles that drift past the planetesimal seeds.  If even a modest fraction are captured and concentrated, then a large core could grow quickly very close to the Sun, as in the present model.

Two recent papers have proposed that the early Solar System contained a system of large terrestrial planets closer-in than Mercury's orbit, and that these were destroyed.  This idea is motivated by the very abundant population of close-in super-Earths and mini-Neptunes discovered by recent exoplanet surveys~\citep{mayor11,howard12}.  While we find the idea that the Solar System once contained additional short-period planets provocative, we disagree with the proposed mechanisms for their destruction, as we explain below. 

\cite{volk15} suggested the inner Solar System once contained a system of planets analogous to the Kepler-11 system~\citep{lissauer11}.  An instability in the planets' orbits triggered a series of giant impacts~\citep[which are indeed common in models for the origin of these planets; ][]{terquem07,raymond08a,hansen12,cossou14,pu15}. If these impacts are energetic enough they would grind the planets to dust, which was subsequently expelled from the system by a combination of radiation pressure, Solar wind drag and Poynting-Robertson drag~\citep[see, e.g.][]{wyatt08}.  However, a cursory examination of the relevant collisional environment suggests that, while the impacts between these planets may indeed have been erosive, they would not have fallen in the ``super-catastrophic'' regime defined by \cite{leinhardt12} and \cite{genda12}.  Several studies have shown that the debris generated by erosive collisions is simply swept up by the surviving planets~\citep{kokubo10,chambers13}.  Thus, we do not expect that a system of close-in terrestrial planets could self-destruct.

\cite{batygin15} proposed that, in the context of the Grand Tack model, the fully-grown Jupiter's migration generated a massive pulse of collisional debris that swept inward and pushed a primordial system of Kepler-11 like planets onto the young Sun.  While we agree that Jupiter may indeed create pulses of debris~\citep[e.g.][]{turrini12}, we see at least three strong arguments against \cite{batygin15}'s idea.  First, accretion models find that collisional debris actually accelerates accretion onto the larger bodies rather than simply drifting inward~\citep{kenyon09,leinhardt09b}.  Second, the mechanism by which 100~m-scale planetesimals ``push'' planets inward arises from the ``superparticle'' approximation and is unlikely to be physical. Simulations usually cannot resolve the billions of objects and thus only include a modest number of superparticles (typically a few thousand), which as an ensemble have the appropriate mass but dissipate as though they were much smaller bodies.  We utilized the same approach in the simulations from section 2.  As small (sub-km-sized) planetesimals drift inward quickly due to aerodynamic gas drag~\citep{adachi76}, they are captured in exterior resonances with the planets they encounter.  If the planetesimals remain in resonance and dissipate efficiently, simulations find that the energy lost from gas drag is effectively taken from the planets' orbits, causing both the planets and planetesimals to spiral inward rapidly~\citep[see][]{thommes03,raymond06b,mandell07}.  However, an abundant population of spatially-confined small planetesimals (because they share the same resonance with a given planet) would certainly self-interact and grow.  Larger bodies must inevitably grow in these exterior resonances with planets, where small planetesimals are concentrated.  Since large bodies dissipate much less efficiently than small ones, the inward push felt by the planet would decrease.  The very mechanism by which a population of small, highly-dissipative objects pushes a planet inward is therefore unlikely to be correct~\citep[see also discussion in][]{mandell07}.  To truly validate or refute this mechanism would require simulations with a code capable of reliably calculating the long-term collisional and dynamical evolution of the in-spiralling planetesimals embedded in the gaseous protoplanetary disk.  Such a code could be built on collisional codes created for debris disk studies~\citep[see, e.g.][]{nesvold13,kral13}.   Third, models of protoplanetary disks suggest that, at least in their later phases, stars accrete magnetically~\citep[see review by][]{bouvier07}.  Models of magnetized accretion suggest that protoplanetary disks have inner cavities~\citep[e.g.][]{romanova03}.  It is generally thought that the inner edge of disks corresponds to the corotation distance, where the orbital period matches the star's rotation~\citep{armitage11}. Planets cannot simply be pushed onto the stellar surface to be destroyed, as invoked by \cite{batygin15}.  Rather, planets are either retained at the inner edge, where there is a strong positive type I torque~\citep{masset06}, or scattered into the cavity itself, where gasdynamical effects cease or are at least severely weakened.  Indeed, the migration model for the origin of close-in exoplanets has shown how inward migrating planets naturally pile up at the inner edge before a later phase of destabilization~\citep{terquem07,cossou14}.  We also find it ironic that \cite{batygin15} invoke the rapid inward drift of solids to destroy super-Earths whereas many other models invoke it to create them~\citep{boley13,chatterjee14}.

Finally, \cite{izidoro15a} proposed that the Solar System did not form close-in super-Earths because Jupiter formed quickly and blocked their inward migration.  In their model, the building blocks of Saturn's core and the ice giants would have migrated inward to become super-Earths but were held back by Jupiter~\citep[this model can indeed reproduce the masses and orbits of the ice giants; see ][]{izidoro15b}.  However, this model simply does not address the Solar System's terrestrial mass deficit interior to Venus, and another mechanism is required to understand the lack of very close-in planets.

\subsection{Implications of Jupiter's outward-migrating core for Solar System formation}

We have proposed a new scenario for the origin of Jupiter's core, motivated by the models of \cite{chatterjee14} and \cite{li15}.  To summarize, we suggest that Jupiter's core may have accreted rapidly in the innermost parts of the protoplanetary disk, perhaps from the accumulation of inward-drifting pebbles~\citep[as in the model of][]{chatterjee14}.  Once the core reached a few Earth masses it may have started to migrate outward (see Fig.~1).  Additional torques generated by the heat of accretion~\citep{benitez15} and the dynamics of the corotation region~\citep{paardekooper14,pierens15} may have helped sustain outward migration in the face of other effects that favor inward migration~\citep[e.g.,][]{cossou13}.  If the core's outward migration occurred on a $\sim 10^5$ year timescale then it could have cleared out the inner $\sim 0.4-1$~AU and could potentially explain the absence of terrestrial planets interior to Mercury (see Fig.~2.).  Jupiter's migrating core would have accreted a modest amount of mass and it may even have stimulated the growth of Saturn's core (see Figs.~3 and 5).

Of course, the story does not end when Jupiter's core reached several AU.  Presumably, it then accreted a massive gaseous envelope and became a gas giant~\citep{pollack96} and should thus have started to type II migrate inward~\citep{lin86}. The large Earth/Mars mass ratio indicates a mass depletion in the Mars region during terrestrial accretion~\citep{hansen09,raymond09c,izidoro15c}.  Our model is consistent with the later evolution proposed in the Grand Tack~\citep{walsh11} and classical models of terrestrial planet formation~\citep[reviewed in][]{raymond14}.

The migration of gas giant core(s) from the innermost regions of the gas disk through the terrestrial region of the disk have strong implications for the terrestrial planets. The giant planet core forms at a special location in the disk, presumably a pressure bump created by the inner edge of the MRI-inactive dead zone~\citep{chatterjee14} but possibly a series of opacity transitions due to sublimation fronts of various silicate species. Growing planetesimals and embryos in the terrestrial region do not have this advantage so they are expected to be smaller~\citep[see for instance][]{morby15a}. Thus, their primary growth mechanism is likely pebble accretion~\citep{lambrechts12}. When the migrating core passes through the region, during the shepherding and scattering process, it excites their eccentricities and inclinations. Increased inclinations strongly inhibit pebble accretion because the disk of pebbles is very thin~\citep{levison15}, and larger eccentricities also inhibit pebble accretion because it increases the relative velocity between the embryo and pebbles. Thus, embryos scattered interior by Jupiter would initially have their growth stalled. Contrast this to embryos that remain always exterior to the migrating core. If these embryos are widely spaced enough due to inefficient planetesimal formation, then they grow by pebble accretion much more efficiently~\citep{kretke14}.

From these considerations, we can speculatively propose a grand view of planet formation in the Solar System, in which we see four pairs of planets. The cores of the gas giants, Jupiter and Saturn, migrated outwards together from near the inner edge of the gas disk. Jupiter was the original core and shepherded Saturn's core -- which was less massive -- until Jupiter arrived at the zero torque boundary (see Fig.~1). Pebble accretion is a runaway process~\citep{lambrechts12}, and so Jupiter's core grew faster and began gas accretion earlier than Saturn's core. This timing delay is exactly what the Grand Tack hypothesizes is necessary for the first inward and then outward migration of Jupiter and Saturn~\citep{walsh11,pierens11}. In this model, Jupiter and Saturn were ultimately responsible for both the inner and outer edges of the terrestrial region.  While this model is attractive, we stress that the outward migration of Jupiter's core is not explicitly linked with the Grand Tack model.

In the outer Solar System, embryos could grow efficiently by pebble accretion~\citep{lambrechts14} to several Earth-masses as long as their eccentricities and inclinations remained small. There were likely several large cores at this stage, representing the precursors of Uranus and Neptune. Giant collisions between these cores -- whose migration was held back by Jupiter~\citep{izidoro15a} -- would naturally reproduce the obliquities~\citep{morby12c}, masses and orbits of the ice giants~\citep{izidoro15b}.  This accretion often produces extra embryos~\citep{izidoro15b}, such as the hypothesized fifth giant planet necessary to scatter Jupiter and avoid unfortunate secular resonances in the terrestrial region during a Nice model giant planet instability~\citep{nesvorny12,brasser13,kaib16}. Scattering of leftover cores during this stage is also a natural way to explain the wide orbit of the putative ``Planet 9''~\citep{batygin16}.

Regarding the terrestrial planets, Mercury and Mars are likely leftover embryos. They represent the initial size distribution of embryos before Jupiter's core migrated through the terrestrial region, with modest later growth due to inefficient pebble accretion as their orbits dynamically cooled. Considering the Grand Tack, Venus and Earth are then the products of the mergers of a number of Mercury to Mars mass embryos, while Mars is an embryo scattered off the outer edge of the terrestrial disk as Jupiter migrated back to $\sim$5 AU~\citep{walsh11}. Mercury may also be an embryo scattered off the inner edge of the terrestrial disk although this occurs only rarely~\citep{jacobson14b}.  Another possibility is that Mercury represents the first embryo scattered inward by Jupiter's outward migrating core. Mercury could thus have become isolated from the rest of growing terrestrial planets. To speculate further, as Jupiter's migrating core migrated outward it would have excited the eccentricities of any planetesimals or embryos in Mercury's vicinity, simply because those objects left behind were necessarily scattered.  This could potentially have created the conditions for the erosive collisions needed to explain Mercury's peculiarly high metal to silicate ratio~\citep{benz07,asphaug14}. To be viable, this model requires collisional debris to have been rapidly removed from the system due to high velocity planetesimal-planetesimal collisions~\citep[e.g.,][]{wyatt08,carter15}.  To match the present-day Mercury roughly half of the initial mass of an Earth-composition embryo would have needed to be removed.  While this is indeed a challenge to explain dynamically, it is a much less stringent constraint than the 100\% mass removal required by \cite{volk15} to completely destroy super-Earths.  Continuing along this line of thinking, since orbital velocities are lower further from the Sun, erosive collisions no longer substantially change iron to silicate ratios but continue to evolve Mg/Si ratios by preferentially removing crust, thus potentially explaining the achondritic composition of the Earth~\citep[e.g.][]{campbell12}.

\begin{figure}
  \begin{center} \leavevmode 
 \epsfxsize=8.75cm\epsfbox{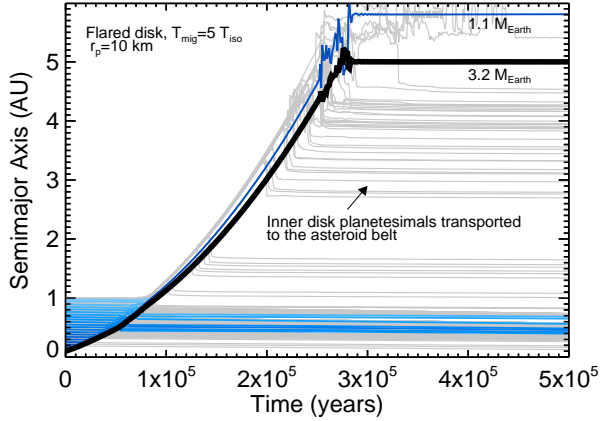}  
    \caption[]{A simulation starting from an annulus of material in which asteroids were transported from the terrestrial planet-forming region to the asteroid belt.  The inner parts of the disk were cleared and a $1.12 \mearth$ core survived at 5.8 AU, in 4:5 resonance with Jupiter's core. Embryos are shown in blue and planetesimals in gray. In this case, $\tau_{mig} = 5 \tau_{iso}$, $r_p$ = 10~km and the disk was flared. } 
     \label{fig:at_ast}
    \end{center}
\end{figure}

It has been proposed that the parent bodies of iron meteorites originated in the terrestrial planet forming region and were implanted into the asteroid belt~\citep{bottke06}.  The mechanism proposed by \cite{bottke06} requires a continuous chain of embryos spanning the inner Solar System, which pass planetesimals outward via gravitational scattering.  This is at odds with the constraint of a strong mass deficit beyond 1 AU imposed by Mars' small mass.  However, Jupiter's outward-migrating core provides a natural mechanism for the outward transport of planetesimals from the very inner parts of the disk.  Planetesimals that are shepherded by the migrating core can be destabilized by gravitational interactions with shepherded or resident embryos, and then be left behind.  Figure~\ref{fig:at_ast} shows an example simulation in which a large number of planetesimals was deposited in the asteroid belt as Jupiter's core migrated through.\footnote{Note that the system of surviving terrestrial embryos in Figure 5 is slowly being pushed inward by the mechanism proposed by \cite{batygin15}. Planetesimals trapped in exterior resonances with the embryos are dissipating energy due to gas drag and pushing the embryos inward.  We reiterate that we do not consider this a true physical mechanism but rather one that is created by the numerical limitations of our simulations.}  Given the much stronger damping felt by planetesimals due to aerodynamic gas drag, planetesimals may be shepherded in a much wider range of resonances than embryos~\citep{mandell07}. This means that embryos and planetesimals are often spatially separated during shepherding, as can be seen during the core's migration in Fig.~5, where many shepherded planetesimals' orbits are exterior to the shepherded embryo's.  Implantation of terrestrial planetesimals into the asteroid belt -- which can be seen to a smaller degree in Fig.~3 -- is common in our simulations. This mechanism operates in both of our sets of simulations, but is more efficient in simulations with an edge in the initial distribution of embryos and planetesimals at 1 AU.  As for the case of additional shepherded cores, the presence of embryos from 1 to 5 AU acts to destabilize shepherded planetesimals such that only a small fraction of terrestrial planetesimals are transported to the asteroid region.

%Planetesimals that are shepherded by the migrating core are protected from shepherded embryos because, due to the much stronger damping felt by planetesimals due to aerodynamic gas drag, planetesimals may be shepherded in a much wider range of resonances than embryos. This means that embryos and planetesimals are often spatially separated during shepherding 

\subsection{Global view of outward-migrating cores}

Might all giant planet cores form in the innermost parts of their protoplanetary disks and migrate outward~\citep[as proposed by][]{li15}?  To address this question, we first must understand whether the outward migration of cores is a generic process or confined to only a limited range of conditions.  This is determined by the properties and evolution of gaseous protoplanetary disks.

The migration speed and direction of planetary cores is determined by the underlying disk structure~\citep{paardekooper11}. As the disk evolves in time its structure changes and with it the zones of outward migration, which are in many cases caused by transitions in opacity at the silicate and water ice lines~\citep[see Fig.~1;][]{bitsch15}. Clearly the migration maps presented in Fig.~1 support our idea of outward migration of planetary cores from the inner edge of the disk to a few AU. However Fig.~1 only considers one set of disk structure parameters ($\alpha$, $\nu$, Z, etc.). Is this pattern of outward migration just available for a small subset of disk parameters or is it a general outcome and might all giant planet cores form in the very inner parts of the protoplanetary disk and migrate outwards?

\begin{figure}
%  \begin{center} 
%  \leavevmode 
 \epsfxsize=8.25cm\epsfbox{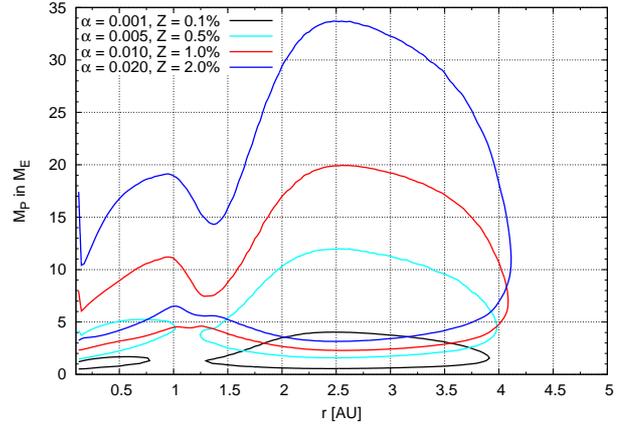}  
% \centering
%\includegraphics[width=\columnwidth]{MdotSean.pdf}
%\includegraphics[scale=0.5]{MdotSean.pdf}
 \caption{Regions of outward migration in disks with an accretion rate of $\dot{M} = 1 \times 10^{-8} M_\odot/yr$, but with different $\alpha$ values and metallicities. If a planet is inside the solid lines, it will migrate outwards. If a planet is outside the solid lines, it will migrate inwards.  Outward-migrating planets are expected to converge at the right side of the enclosed contours, at convergent zero-torque zones.  }
   \label{fig:migcont}
%       \end{center}
\end{figure}

In Fig.~\ref{fig:migcont} we display the zero-torque lines in disks with an accretion rate of $\dot{M} = 1 \times 10^{-8} M_\odot/yr$, but with different $\alpha$ values and metallicities. The viscosity, surface density and metallicity in a disk can be exchanged so that the disk has the same thermal structure and density gradients (see section 5 in \cite{bitsch14} for a more detailed explanation). Nevertheless the migration behavior of embedded planets is different, because the saturation of the corotation torque does not only depend on the radial gradients of surface density, temperature and entropy, but also on the thermal conductivity and viscosity of the gas~\citep{paardekooper11}. Effectively the radial extent of the regions of outward migration is the same for all values of viscosity and metallicity, because the radial extend of the regions of outward migration is determined by the strength of the radial gradients in the disk. However the disk can accompany different planetary masses in the region of outward migration for different values of viscosity and metallicity. In particular a large viscosity and metallicity allows more massive planets to migrate outwards compared to low viscosities and low metallicities.

The effect of metallicity is to increase the opacity and thus heat the disk.  In a hotter disk, the steep gradients needed for outward migration shift outwards and with it the regions of outward migration~\citep{bitsch15}. Higher-mass planets can migrate outward in higher-metallicity disks. The disk's viscosity has the same general effect.  At fixed disk mass, a disk with higher viscosity is hotter. The regions of outward migration are therefore farther from the central star.  Higher-mass planets also migrate outward in higher-viscosity disks.  

Outward migration of cores from the inner disk to a few AU appears possible for a wide range of disk parameters.  The exceptions are disks with very low viscosities or very low metallicities.  This implies that our proposed migration scenario may have happened in many systems.

We envision a model for planetary system formation built on rapid growth of close-in cores followed by subsequent outward migration. Large cores grow quickly in the innermost parts of the disk by accreting a fraction of inward-drifting pebbles~\citep[as in][]{chatterjee14,boley14,li15}.\footnote{In the proposed model the cores' growth happens close-in from pebbles that condensed across a broad radial swath of the disk~\citep[see][]{lambrechts14}.  This model should not be confused with the strict ``in-situ accretion'' model, which conjectures that hot super-Earths form locally from locally-condensed solids~\citep[proposed and subsequently rejected by][then re-proposed by \cite{chiang13}]{raymond08a}.  In-situ accretion requires extremely massive disks very close to their stars~\citep{chiang13}. There are many arguments against the strict in-situ accretion model~\citep[see][]{raymond08a,raymond14,schlichting14,schlaufman14,raymond14b,inamdar15,ogihara15}.  We propose that models that invoke the inward drift of solids followed by accretion at close-in orbital radii should be referred to as a separate category of ``drift'' models rather than being lumped together with in-situ accretion.} Once cores reach a critical mass of a few $\mearth$ (depending on the local disk conditions), they enter a region of outward migration (see Figs.~1 and 6).  Cores migrate out to zero-torque zones, which can be located anywhere between $\le 1$~AU and $\ge 10$~AU depending on the disk's properties and evolutionary state~\citep{bitsch13,bitsch14,bitsch15}.  The cores subsequent migration is much slower: cores remain at zero-torque zones, which themselves shift inward and to lower masses as the disk evolves~\citep[see for example][]{lyra10,coleman14}. Interactions between growing cores can also affect their migration~\citep{cossou13,pierens13}. Cores may either migrate back inward at later times to form systems of super-Earths or accrete enough gas to become gas giants~\citep[][although their later migration could potentially bring them back inwards -- see \cite{bitsch15b}]{cossou14}.

Consider the embryos left behind by a core's migration.  If these (presumably terrestrial) embryos were roughly the size of Mercury and Mars, then their orbital eccentricities and inclinations were inefficiently damped by both aerodynamic drag (too large) and tidal gas damping~\citep[too small;][]{ida08b}. These embryos' orbits remain excited and do not efficiently accrete pebbles, remaining in place and small throughout the gas disk phase~\citep{lambrechts12,levison15}. However, if the embryos were initially larger then tidal damping would have been stronger. Once their orbits were dynamically damped, embryos would grow efficiently due to pebble accretion and then undergo gas-driven migration~\citep{terquem07,cossou14}.  We imagine that leftover embryos could follow two divergent paths: small embryos would remain small and slowly accrete into terrestrial planets, and large embryos would grow quickly and migrate inward to become super-Earths. The Solar System would thus fall into the first category and the abundant population of systems of hot super-Earths into the second. 

Naturally, this model must match observations to be viable.  Given that super-Earths are several times more abundant than outer gas giants~\citep{mayor11,howard12}, most cores must migrate back inward to become systems of close-in super-Earths. In addition, to match the strong giant exoplanet-metallicity correlation~\citep{gonzalez97,fischer05} there must be a mechanism by which cores that are stranded at large orbital distances preferentially orbit metal-rich stars (see Fig~6).  Despite these challenges we consider this a promising idea.

Finally, we wonder: why would the Solar System only have formed a single core in the innermost parts of the disk?  Given the large amount of mass available in the disk, and the inevitability of pebble drift in the context of this model, it would seem natural that many cores could form.  One possibility is that Jupiter's core reached the ``pebble isolation mass,'' carved a gap in the local pebble field and thus acted as a barrier to the inward flux of pebbles from farther out.  In that way, Jupiter's core could have migrated outward and prevented additional cores from forming by the same mechanism.  \cite{lambrechts14b} estimated the pebble isolation mass as $M_{iso} \approx 20 \left(\frac{a}{\rm 5 AU}\right)^{3/4} \mearth$, which gives $M_{iso} \approx 6 \mearth$ at 1 AU (note that this value depends on the disk parameters, in particular the disk's vertical scale height).  A $\sim 3 \mearth$ core may not have completely blocked the pebble flux -- especially as it migrated outward -- but it would likely still have thinned the flux significantly in the inner disk, and to an increasing degree as the core grew further.  If all cores originated close-in, this would predict that systems of hot super-Earths should not coexist with distant gas giants~\citep[this anti-correlation was in fact predicted by][but in the context of Jupiter blocking the inward migration of icy cores]{izidoro15a}.  However, multiple systems containing both super-Earths and more distant gas giants are known, such as Kepler-48~\citep{steffen13,marcy14} and Kepler-90~\citep{cabrera14}.  Of course, the prediction asserted above is complicated by the fact that locally-condensed solids can also grow into super-Earths, the possibility that cores can grow at large orbital radii~\citep{lambrechts14,levison15,bitsch15b} and the fact that Jupiter is not a perfect barrier to their inward migration~\citep{izidoro15a}.  

It is also possible that cores may not form continuously.  If cores grow by gravitational collapse from pebbles that have accumulated at a pressure bump~\citep{chatterjee14}, then there may be a critical ``loading'' of pebbles required to trigger instability.  If this degree of loading is not met, pebbles may simply remain trapped at the pressure bump until the gaseous disk dissipates.  The pebbles would then be quickly ground to dust and removed from the system via radiation pressure, stellar wind drag and Poynting-Robertson drag~\citep{wyatt08}.  Perhaps the Solar System's disk only lasted long enough for a single gravitational instability event to occur; a second collapse may have been further delayed by a decreased pebble flux from filtering by the first (Jupiter's) core.

\section*{Acknowledgements}
We thank referee Aaron Boley for a thorough and helpful report that improved and clarified the paper. We gratefully acknowledge discussions with D. Lin, A. Pierens, A. Crida, A. Morbidelli, E. Lega, J. Szulagyi, and M. Lambrechts.  S.N.R., A.I. and B.B. thank the Agence Nationale pour la Recherche for support via grant ANR-13-BS05-0003-002 (project {\it MOJO}).  B.B. thanks the Knut and Alice Wallenberg Foundation and the Royal Physiographic Society for their financial support. S.A.J. was supported by the European Research Council (ERC) Advanced Grant ``ACCRETE'' (contract number 290568).  We also thank Avi Mandell for his contributions to an earlier version of the code used here.  

This paper is dedicated to SNR's wife Marisa and to the partners of all astronomers whose migration makes their research possible.

%\bibliographystyle{mn2e_b}
%\bibliography{refs.bib}

\end{document}